# Numerical generation of random fiber bundles and the influence of microstructural properties on mechanical behavior

Xinling Song[1*], Gilles Hivet[2], Audrey Hivet[2], Anwar Shanwan[2]

1: Research Institute of Science and Technology Innovation, Civil Aviation University of China, Tianjin, China.

2: LaMé, University of Orléans, Orléans, France.

E-mail: songxl0815@outlook.com, gilles.hivet@univ-orleans.fr, audrey.hivet@univ-orleans.fr, anwar.shanwan@univ-orleans.fr

**Abstract**

Understanding the mechanical behavior of quasi-parallel fiber networks is essential for improving the manufacturing processes of fiber-reinforced composites. Mesoscale models of dry yarns and reinforcements require constitutive laws that accurately reflect the heterogeneous microstructure of fiber bundles. This study aims to develop a numerical generator of random fiber bundles for microscopic parametric studies of compaction behavior. A real fiber bundle was first reconstructed from X-ray microtomography data, and the numerical strategy was validated by tracking fiber cross-sections along the bundle length, with a fiber-position error of 5.2%. Based on this validated framework, an experiment-independent generator was established to create parameterized fiber bundles. The generated bundles reproduced the experimental compaction response with good agreement. Parametric results showed that increasing fiber waviness enhances inter-fiber interactions, increases transverse stiffness, and requires a higher load to reach the same fiber volume fraction. This framework provides a useful microscopic basis for studying fiber-bundle deformation mechanisms and for developing future mesoscopic constitutive laws.

## 1. Introduction

Fibrous reinforcements are used in various industries such as aeronautics or the automotive sector [1-3], because they enable substantial gains in the weight/performance ratio [4-6]. During the different manufacturing processes such as Resin Transfer Molding, the fibrous structure are subjected to various mechanical loads, including compression, tension, bending [7-9]. These loading conditions may induce deformation and damage in the fabric [10, 11], thereby affecting the quality of the final part. Numerical simulation therefore provides a cost-effective and efficient way to investigate the mechanical behavior of fibrous media.

Mesoscale approaches have been widely developed to model the mechanical response of dry or prepreg textile reinforcements [12, 13]. Some researchers have further extended these methods to micro-meso-macro frameworks in order to relate structural features across scales to the macroscopic response of fabrics. For example, Sasa Gao et al. adopted the fiber/martix as the representative volume element (RVE) to investigate the influence of inter-yarn shear angle on the impact damage behavior of twill carbon fiber woven laminates [14]. Bouhala et al. developed the micro-to-macro model of woven composites using TexGen and validated it by three-point bending tests [15]. Although these approaches have significantly improved the prediction of fabric-scale behavior, they still rely on homogenized descriptions at the bundle level and therefore cannot explicitly represent fiber-scale interactions and rearrangements within bundles.

Because the mechanical response of a fibrous medium is governed not only by the fibers themselves but also by their local interactions, several studies have moved toward microscale modeling of numerical fiber bundles. Lin et al. used a three-phase model of fiber-interface-matrix to analyze the influence of fiber orientation on the interface penetration threshold, and the effective volume fraction of the interface, which assumes ideal straight fibers and uniform interfaces [16]. Wang et al. developed a random fiber distribution generation method for the adaptive fiber shaking module by combining the Delaunay triangulation algorithm, which can generate fiber bundles with randomly distributed fiber positions [17]. Kumar et al. created random bending and straight fiber units and characterized them based on straightness, fiber spread, and fiber orientation distribution. They concluded that the initial fiber velocity parameter has the greatest impact on fiber mechanics [18]. Pham et al. focused on the transverse behavior of a KM2 yarn at microscopic and homogenized scale. The results showed the reliability of creating fiber bundles and compaction was verified by comparing it with a homogenized model [19]. However, these studies still largely rely on idealized geometries or simplified fiber distributions, which limits their ability to represent the intrinsic heterogeneity of real fiber bundles.

Obtaining realistic bundle geometry is therefore a key prerequisite for accurate mechanical modeling. X-ray computed tomography (XCT) provides direct three-dimensional observations and quantitative information, making it a valuable tool for analyzing deformation mechanisms under different loading conditions [20-22]. XCT was used by Melenka et al. [23], Toda et al. [24], Henyš et al. [25], and Haji et al. [26] to characterize the fiber bundle structure and extract the skeletonized structure of the samples. Bral et al. constructed a numerical model based on the geometric model of real fiber bundles obtained by CT scanning, verified the model through tensile and bending simulations and experiments, and discussed that yarn twist can improve fiber bundle stiffness [27]. Zhang et al. proposed a model based on CT images, obtained in-situ fiber orientation, length and continuity data of fiber bundles, and used two datasets of different lengths to study the mechanical behavior of ring-spun yarn [28]. These studies have laid an important foundation for the characterization and reconstruction of fibrous media. Nevertheless, most of them remain focused on reconstructing existing geometries from images. Numerical fiber bundles with controllable microstructural features, such as fiber misalignment and waviness, are still rarely generated for systematic parametric analyses, and the influence of these features on mechanical behavior has not yet been fully clarified.

The aim of the present work is to improve the understanding and modeling of fiber-bundle mechanical behavior as a basis for yarn-scale analysis. To this end, a numerical representation of a real fiber bundle is first established together with an appropriate simulation strategy, building on the work of Haji [26]. Microstructural descriptors and the associated analysis tools are then defined to validate both the initial and deformed bundle configurations. Finally, this numerical strategy is extended to a virtual fiber-bundle generator capable of creating stochastic yet near-realistic bundles with controlled parameters, in order to systematically investigate the influence of bundle microstructure on mechanical behavior at the microscopic scale.

## 2 Material and methods

### 2.1 Materials

For this study, two model media were considered, each with 40 quasi-parallel polyester fibers without matrix. Sample 0 was originally developed by Haji et al. [26], whereas Sample 1 was designed in the present work. The mechanical and geometrical parameters of the fibers used in both samples are given in Tab.1.

Table 1. Mechanical and geometrical parameters of each fiber in the fiber bundle.

| Parameter | Value |
|---|---|
| Density ($t.mm^{-3}$) | $1.38e^{-9}$ |
| Young modulus (GPa) | 6 |
| Poisson ratio | 0.25 |
| Friction coefficient [26] | 0.2 |
| Diameter (mm) | 0.5 |
| Length (mm) | 14.5 |

### 2.2 Experimental compaction with XCT

The samples were placed in a rectangular compaction channel/support measuring $16 \times 5 \times 3$ mm and then deformed with a confined compaction n test combined with X-ray tomography, with 15μm of sensitivity for sample 0 and 9μm of sensitivity for sample 1, which were selected as appropriate spatial resolutions with respect to the fiber diameter and the XCT acquisition geometry. As shown in Fig. 1(a), the micro-compaction tool has two parts. The upper part of the support, referred to as the upper jaw, was fixed and connected to a load cell with a capacity of 500 N, to record the compaction load. In parallel, uniaxial compression was applied through the displacement of the lower jaw at a constant speed $V = 1$ mm/min. After positioning the sample, the compaction load was applied incrementally, and X-ray scans were acquired at different compaction steps, with a pixel size $v =$ 9 μm, tube power of 12 W, tube voltage of 60 KV and tube intensity of 200 μA. Fig. 1 (b) shows the load curve obtained for sample 1 with 14 compaction steps.

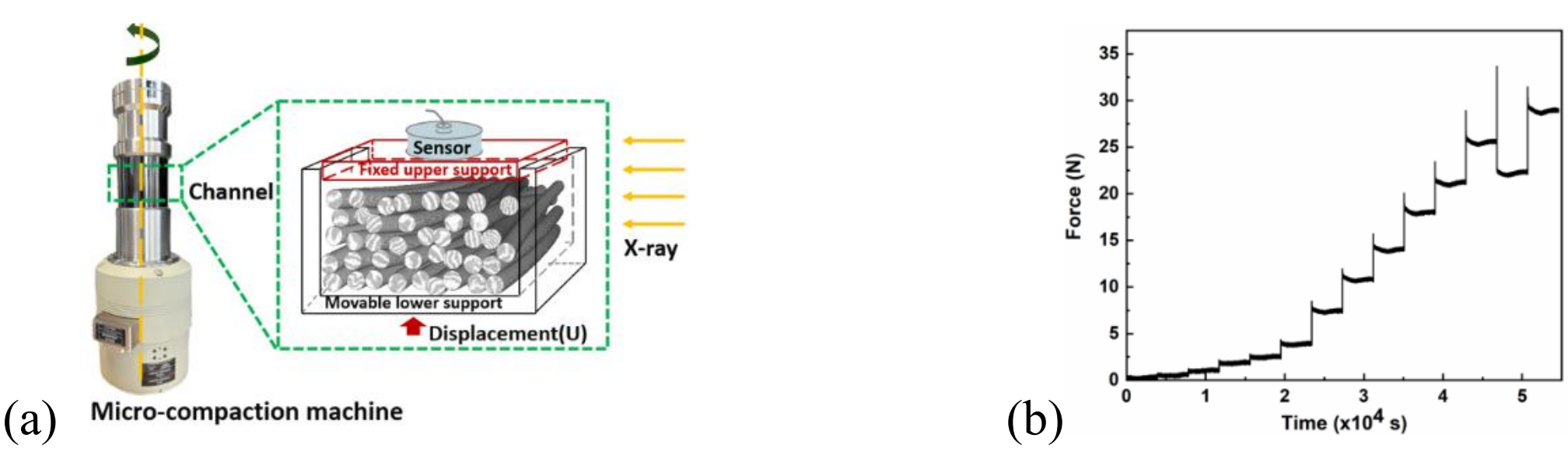

Figure 1. Experimental process. (a) Experimental tool: micro-compaction test combined with X-Ray tomography of the fiber bundle; (b) Load curve for the 14 compaction steps of sample 1.

## 2.3 Numerical reconstruction and compaction simulation

A stack of tomography slices is obtained after each compaction step: the fiber bundle consists of $\mathrm{n}$ slices (Fig. 2(a)) in the fiber length direction, with $\mathrm{n} = \mathrm{L}/v$, $\mathrm{L}$ being the fiber bundle length and $v$ is the tomographic resolution. Here, the fiber section is considered as a circle, and the center points of each fiber are connected to reconstruct the numerical fiber bundle geometry with FIJI and MATLAB, as shown in Fig. 2(b). In this process, the fiber cross-section is detected as a regular circle firstly, then the center point of the circle is calculated by the Circle Hough Transform (CHT) method.

Once the geometrical model of the reconstructed bundle has been obtained, it can be exported in ABAQUS®. In addition, the trays were home-made and presented some flatness and parallelism defects. They were therefore also reconstructed with FIJI and CATIA V5®, as shown in Fig. 2(c). This reconstruction process resulted in a numerical geometry model that closely represents the real fiber bundle and the trays. The numerical compaction test of the fiber bundle was executed in ABAQUS®/EXPLICIT. Linear beam elements B31 were used, as justified with Dr. Haji [26]. The beam element size was $0.3\ \mathrm{mm}$, and their diameter was considered as constant and the same as that of the fibers ($0.5\ \mathrm{mm}$). The compaction channel/support was meshed with discrete rigid R3D4 elements with a size of $0.2\ \mathrm{mm}$, which were validated by a mesh sensitivity test. The material coefficients are those of Table 1. A damping coefficient $\alpha = 10^5$ and compaction speed of simulation $\mathrm{V_{numer}} = 10^3\ mm/min$ were implemented, in order to optimize the calculation time and avoid the inertia effect of the fibers. The contacts were implemented thanks to a Hertz contact law developed by Haji et al. The compaction loads were applied in steps as experiments, as in Fig. 1.

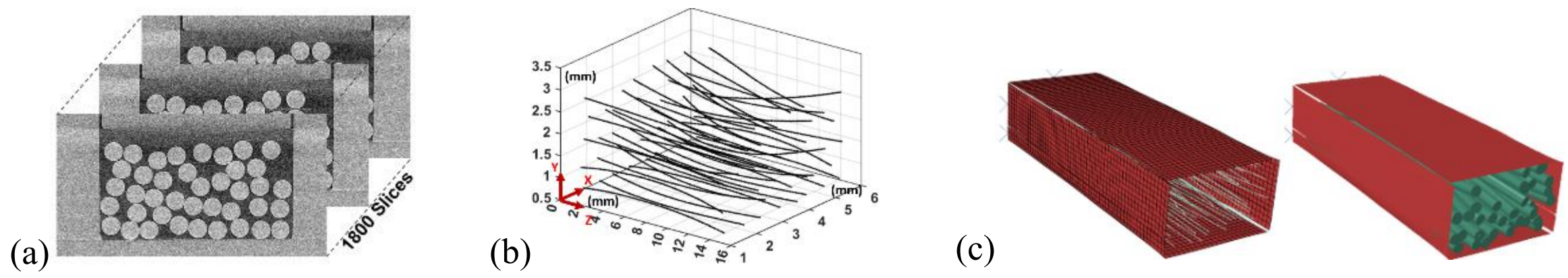

Figure 2. Reconstruction of fiber bundle geometry. (a) Initial tomography slices, (b) Numerical reconstructed centerlines of the fiber bundle, (c) Geometry reconstructed in ABAQUS® (fibers shown in beams/3D rendering (only visual)).

## 2.4 Microstructural descriptors and analysis methods

The strategy presented on Fig. 3 was extended to calculate two types of contact angles between fibers can be distinguished: the contact angle between the fibers in the transverse plane (fiber cross sections) (θ), and the contact angle between the fibers in the length direction (γ). During the compaction of fiber bundles, the fibers transmit contact loads to each other and $\theta$ changes as the fibers move. Also, once a certain amount of contact has been reached, the fibers deform, causing them to continue to move and $\gamma$ to change. Thus, the contact angles are also interesting indicators of the microstructure and its evolution since they are of paramount importance in the load transmission.

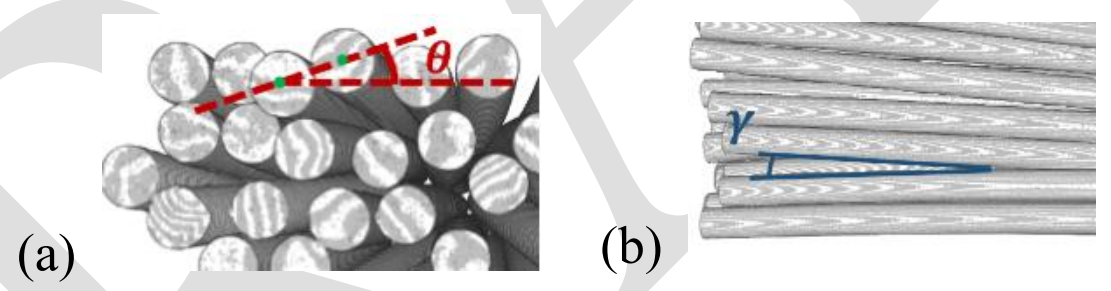

Figure 3. Contact indicators. (a) Contact angle between two fibers in the section (θ), (b) Contact angle between two fibers in the fiber length direction (γ).

Compaction changes the microstructure of the fiber bundle, not only through changes in inter-fiber contact, but also through changes in fiber waviness and local bending. In the present study, waviness is used as the general term to describe the non-straight morphology of a fiber. Its global magnitude is quantified by crimp, whereas its local bending is characterized by the projected curvature in the XZ and YZ planes (Fig. 4). In addition, the spatial distribution of fiber directions is quantified by the second-order orientation tensor.

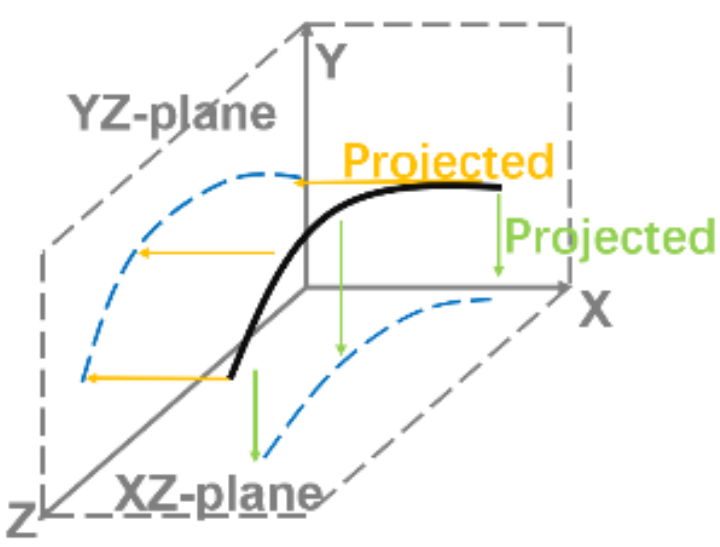

Figure 4. The strategy of projected curvature in the XZ-plane and YZ-plane.

In this case, the crimp of a single fiber is considered as the degree to which the non-straight fibers deviate from linearity:

$$\text{Crimp (\%)} = \frac{L_c - L_0}{L_0} \times 100\% \qquad \text{(Eq.1)}$$

where $L_0$ is the single fiber projected length and $L_c$ is the real length of the fiber. Fig. 4 shows the principle of analysis for the projected curvature of a single fiber in the fiber length direction. It is considered here in both the XZ plane and the YZ plane, in order to characterize the local bending of the fiber path in these two directions during compaction.

In addition, the fiber mis-orientation distribution and orientation tensor were employed to quantify the spatial orientation of the fibers. The second-order orientation tensor can be used to quantify the orientation distribution of fibers, i.e. $A_j$:

$$A_j = \frac{1}{N} \sum_{i=1}^{N} \bar{\mathbf{q}}_i \otimes \bar{\mathbf{q}}_i \qquad \text{(Eq.2)}$$

where $N$ is the total number of fibers, $\bar{\mathbf{q}}_i$ is the orientation tensor of a single fiber $i$ in the compaction step $j$.

## 3 Results and Discussion

### 3.1 Reconstruction quality and validation of the numerical estimator

The pixel size used during the XCT experiment is one of the main parameters to be considered since it enables a more or less accurate description of the fiber section and position. A lower spatial resolution may result in the loss of surface details of the fibers, especially in the contact region between the fibers. To represent the relative position of the fibers accurately, a strategy of contact detection was developed, as shown in Fig. 5. The contact forces transmitted by the contact points between fibers significantly affect the compaction behavior and consequently the fiber motion. The different contact cases are depicted in Fig. 5. The contact is studied by considering two fibers in the slice. The contact detected by the program can be divided into two cases; for each one, the center shift can lead to an inconsistent determination of the contact reality:

- The two fibers are in contact in the real situation (Fig. 5 (a)). Three cases can occur: the first and simplest one occurs when the position of the circle center is on the pixel coordinate point (case (a-3)); in this case, the contact is correctly modeled. Otherwise, if the center position is not on the pixel coordinate point, the center position of the circle will be shifted, causing the two circles to separate (case (a-1)) or intersect (case (a-2)). With this strategy, the error $\varepsilon_1$ made on the numerical center coordinates of the fibers is between $0 < \varepsilon_1 \leqq 0.5$ pixel. These two cases will lead to an inconsistency in the contact determination.
- The two fibers are not in contact in the real situation (Fig. 5 (b)) but the centerlines are very close; in this case, there is a risk that the two circles may come into contact after the circle center is shifted (case (b-1)).

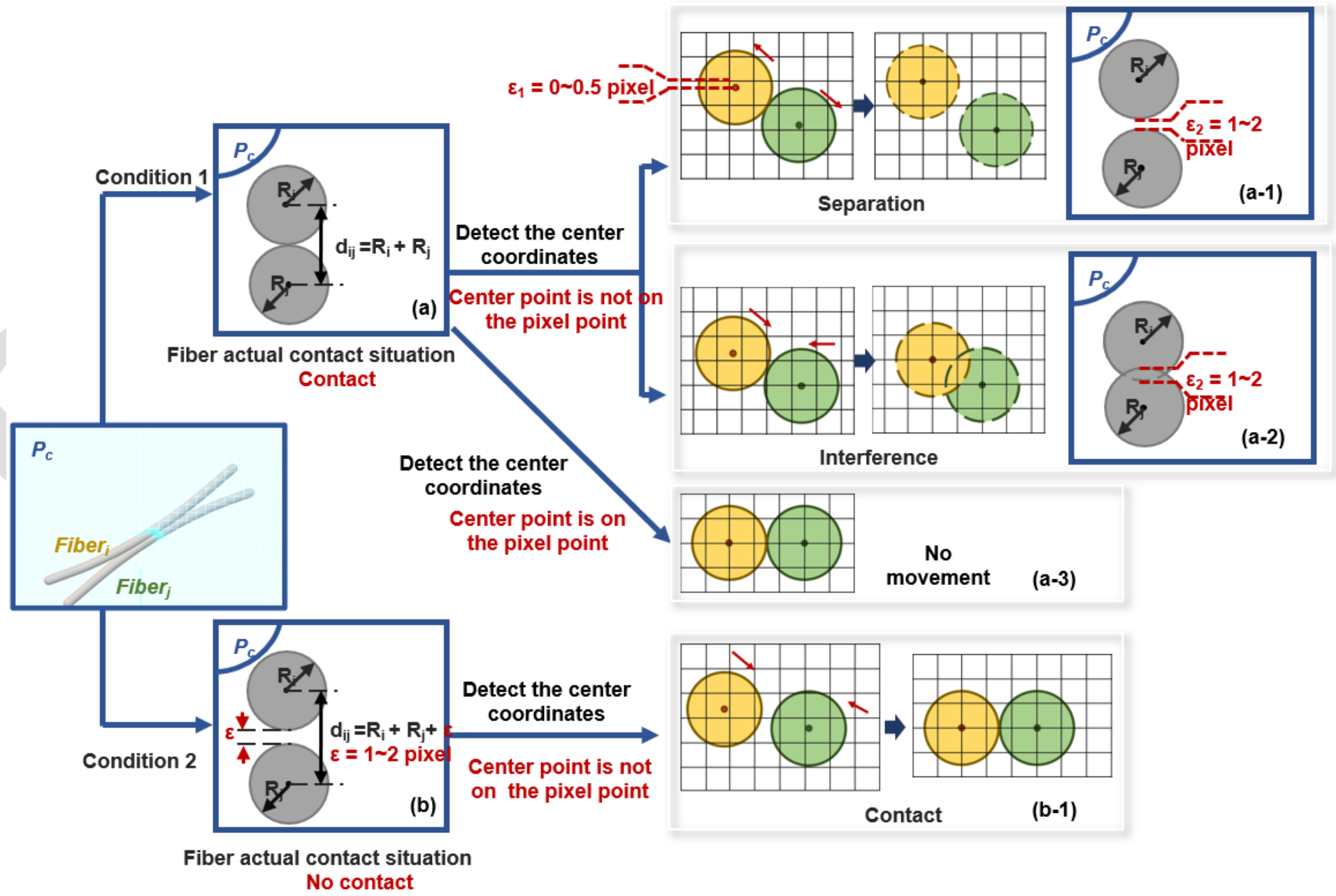

Figure 5. Contact detection.

Inaccuracies in the contact analysis will increase the loss of details introduced by resolution limitations, leading to even more inaccurate predictions of the structure of the fiber bundle after simulation. Finally, in the reconstructed bundle, a gravity step simulation is performed to bring the sample back to static equilibrium. If the model were an absolutely perfect representation of the real bundle, the virtual bundle would be in static equilibrium and the gravity step would not be useful. On the contrary, the farther the model is from the real bundle, the more the fibers will move during the gravity step to reach static equilibrium. That is why the movement of fibers during this gravity step is also a very strong indicator of the quality of the modeling: the less the fibers move during this step, the more accurate the bundle reconstruction is.

After the first gravity step (Fig. 6) imposed on the fiber bundle geometry, static equilibrium was reached, making it possible to quantify the discrepancy between the simulation model and a real bundle. The Fiber Position Deviation Diameter Ratio was calculated ($\mu_{FPDDR}$) as the ratio of the average fiber distance to the fiber diameter for experiments and models (eq. 3 and 4). The results can be seen in Tab.2.

$$\mu_{FPDDR_i} = \frac{\overline{\mathrm{d}_{\mathrm{f_i}}}}{\phi_{\mathrm{f}}} \qquad \text{(Eq.3)}$$

$$\mu_{FPDDR} = \frac{\sum_{\mathrm{i}=1}^{\mathrm{N}} \mu_{FPDDR_i}}{\mathrm{N}} \qquad \text{(Eq.4)}$$

$$\sigma_{FPDDR} = \sqrt{\frac{\sum_{\mathrm{i}=1}^{\mathrm{N}} \left(\mu_{FPDDR_i} - \mu_{FPDDR}\right)^2}{\mathrm{N}}} \qquad \text{(Eq.5)}$$

with $\mathrm{N}$ the number of fibers, $\phi_{\mathrm{f}}$ the fiber diameter and $\overline{\mathrm{d}_{\mathrm{f_i}}}$ the average distance between simulation and experimental fiber position for fiber $\mathrm{i}$.

This clearly illustrates the influence of the voxel size but also shows that the initial modeling obtained for sample 1 ($v = 9\ \mu m$) is accurate since very small changes are observed in the fiber position after the gravity test (5% with a small standard deviation). As mentioned previously, the effect of the gravitational field leads to a micro-adjustment of the internal structure of the fiber bundle, yielding a statically equilibrated structure. In addition, the gravity test indicated that the reconstruction process is powerful.

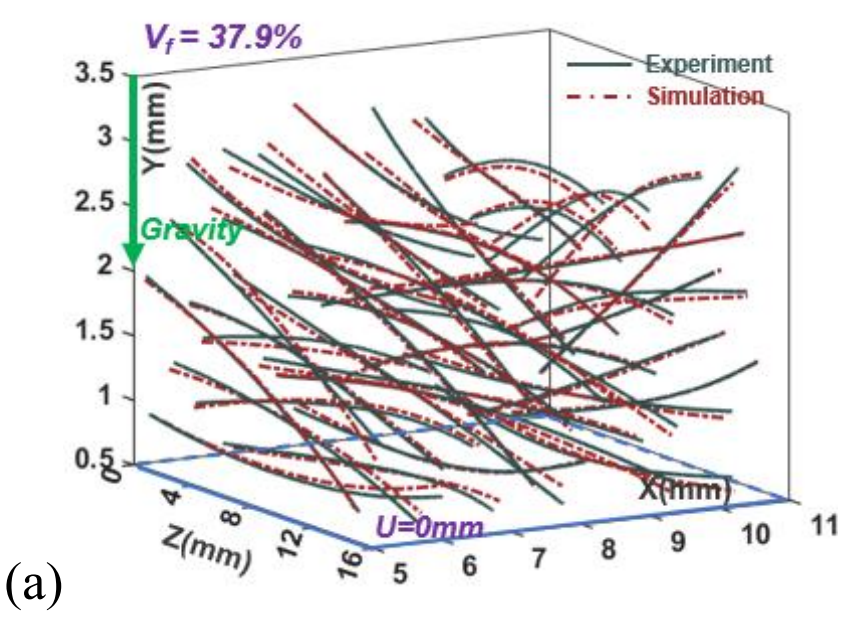

(a)

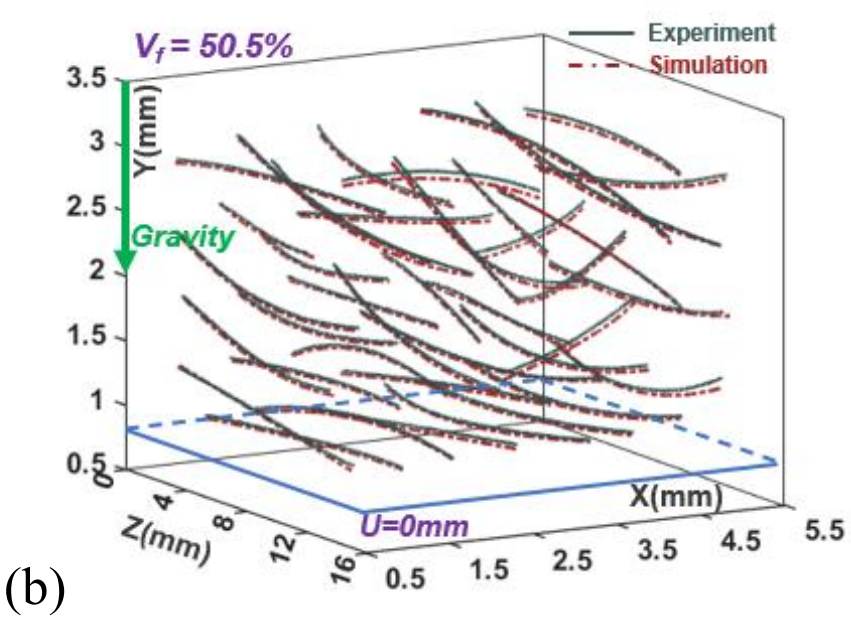

(b)

Figure 6. Fiber bundle centerlines after imposing the gravity field. (a) Sample 0 ($v = 15\ \mu m$), (b) Sample 1($v = 9\ \mu m$).

Table 2. Position deviations of fibers between experiment and simulation (μ: average value; σ: standard deviation value).

| Fiber Position Deviation Diameter Ratio (%) | Sample 0 ($v = 15\ \mu m$) | Sample 1 ($v = 9\ \mu m$) |
|---|---|---|
| $\mu$ | 15.83 | 5.20 |
| $\sigma$ | 5.50 | 1.20 |

To evaluate the performance of the proposed strategy, the confined compaction test was simulated for the numerically reconstructed fiber bundle. Fig. 7 (a) shows the compaction pressure as a function of the volume fraction ($V_f$) of sample 1, and the strain, and Fig. 7 (b) shows the strain ($S_e$), penalty ($P_e$) and kinetic energy ($K_e$) in the compaction process. The simulation curves are in good agreement with the experimental results. Overall, the simulated response is in good agreement with the experimental results. In particular, the pressure drop observed experimentally at step 13 is also captured by the simulation, which strongly supports the consistency of the proposed numerical strategy.

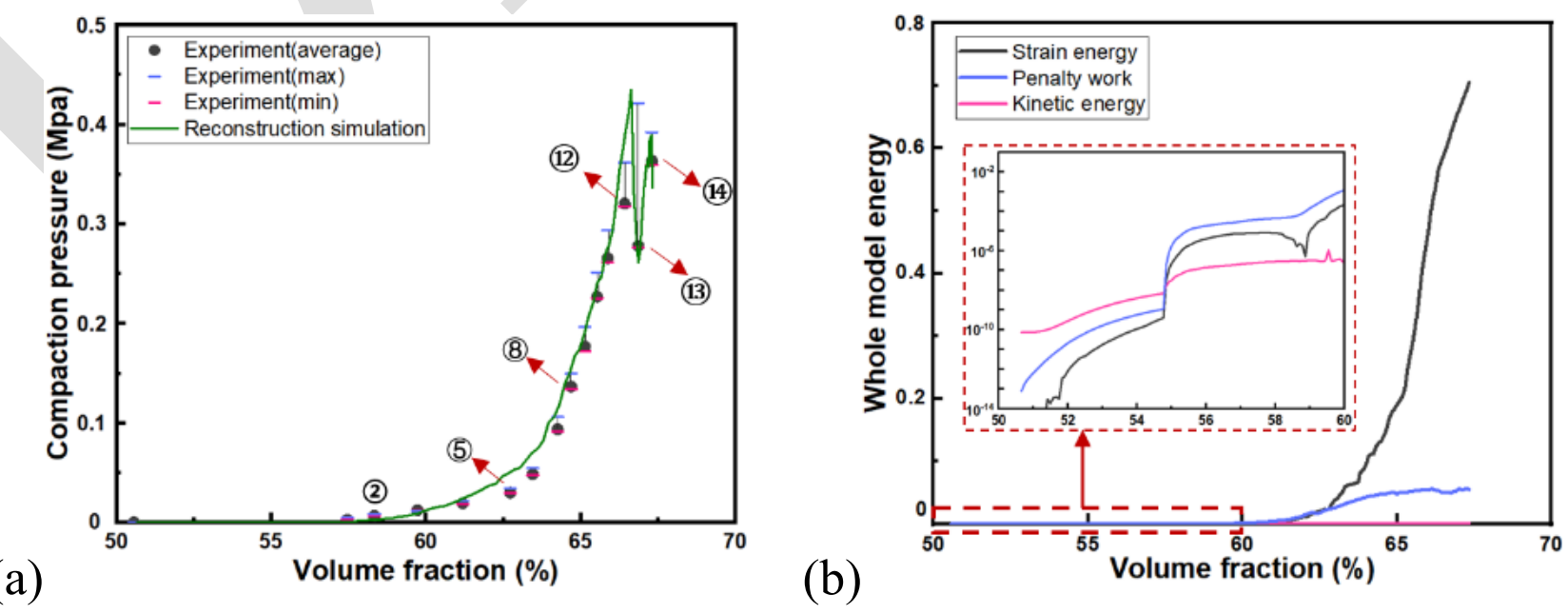

(a) (b)

Figure 7. Compaction simulation for sample 1. (a) Evolution of compaction pressure as a function of volume fraction. (b) Evolution of the energy during compaction.

As shown in Fig. 7 (b), at the initial stage of compaction ($V_f \in [50.5, 55]$ (%)), there are spaces for the fibers to move, so the $S_e$ and $P_e$ of the model are about $10^{-5}$ times of $K_e$. The number of contact points between the fibers increases rapidly with the beginning of the compaction simulation. Then, the contact between the fibers increases when slight movement occurs due to slippage. Around $V_f$ = 55%, the energy of the model is inverted, indicating a shift in the preferred mode of fiber bundle deformation from displacement to fiber deformation. $K_e$ and $P_e$ increase significantly, indicating that the rigid movement of the fibers has reached a threshold. The contact points between the fibers increase, and fiber deformation begins to occur. At $V_f \in [55, 62]$ (%), the contact energy is higher than the kinetic energy of the model, which is due to the fact that during the compaction test the fibers first undergo contact, and then transfer compressive forces through the contact points, thereby deforming the fibers. Beyond $V_f$ = 62%, the response becomes dominated by strain energy, which indicates that further compaction is mainly accommodated by fiber deformation rather than by fiber rearrangement alone.

The fiber centerline positions of Sample 1 after compaction are shown in Fig. 8. As reported in Tab. 3, the average discrepancy between the experimental and simulated fiber positions, normalized by the fiber diameter (0.5 mm), is 14.07% at step 14. Fig. 8 also compares the simulated fiber bundle with the three-dimensional reconstruction obtained from the experiment. Although the overall morphology is well reproduced, local visual deviations remain visible between the two representations. These differences arise from the combined effects of XCT resolution, centerline extraction errors, numerical idealization of the fibers by beam elements with constant diameter, and the gravity-relaxation step introduced to obtain a mechanically admissible equilibrium configuration. In addition, the simulated bundle is displayed from reconstructed centerlines, whereas the experimental configuration is obtained from XCT-based three-dimensional reconstruction, so the two visualizations are not strictly equivalent in appearance. Nevertheless, the global agreement remains satisfactory. The final bundle shape, the compaction response, and the main microstructural descriptors are all reproduced consistently by the numerical model. In particular, both the experimental and numerical compaction curves in Fig. 7 (a) exhibit a pressure drop around step 13 $\left(V_f = 66.9\%\right)$. To clarify the origin of this feature, the microstructural descriptors were analyzed in detail from

steps 12 to 14.

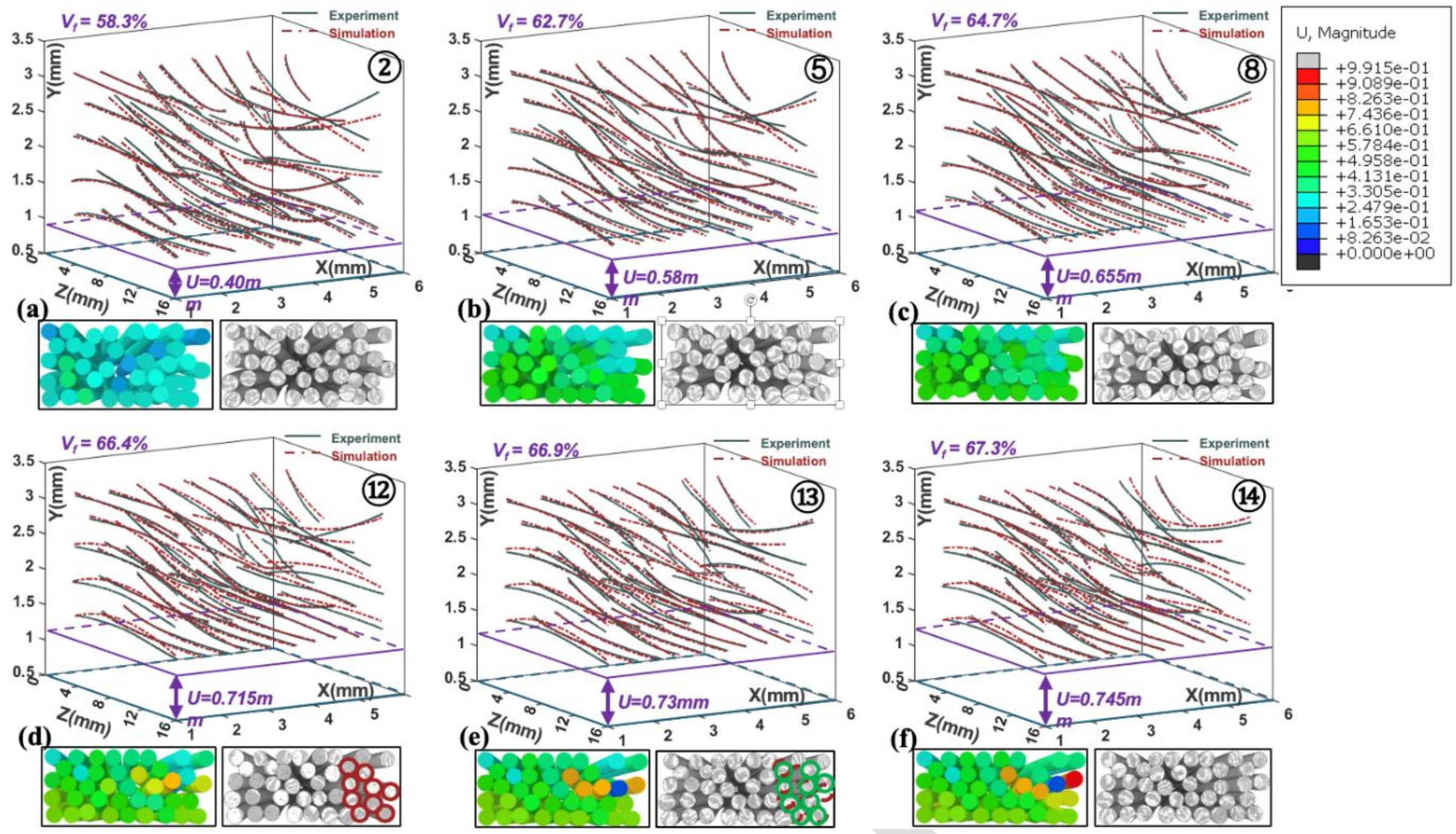

Figure 8. Centerline and 3D rendering of sample 1 after simulation and experiment compaction. (a-f) Step 2, 5, 8, 12-14.

Table 3. Position deviations of fibers between experiment and simulation of sample 1 for compaction steps 5, 12, 13, 14 ($\mu$: average value; $\sigma$: standard deviation).

| Fiber Position Deviation Diameter Ratio (%) | Step 5 | Step 12 | Step 13 | Step 14 |
| --- | --- | --- | --- | --- |
| $\mu$ | 10.95 | 17.00 | 14.48 | 14.07 |
| $\sigma$ | 2.74 | 9.70 | 7.76 | 7.89 |

### 3.2 Microstructural evolution during compaction

Contact is defined as one contact between two fibers in one slice or between one fiber and one plate in one slice, as illustrated on Fig. 9 (a). The total contact number is then obtained by summing all contacts over all slices. Fig. 9 (b) shows the evolution of this indicator for steps 12, 13 and 14. These results show good correlations between the numerical and experimental contact numbers. Fig. 10 presents the distribution of the contact angles for the experimental and numerical sample 1; the results again show a good correlation between experiments and simulations.

Between steps 12 and 13, a slight decrease in the total contact number is observed, while the corresponding contact-angle distributions remain nearly unchanged. This result suggests that the pressure drop is not associated with a large-scale reorganization of the bundle. Instead, it points to a local rearrangement event, in which a limited number of fibers slip and modify both their contact locations and their contact number.

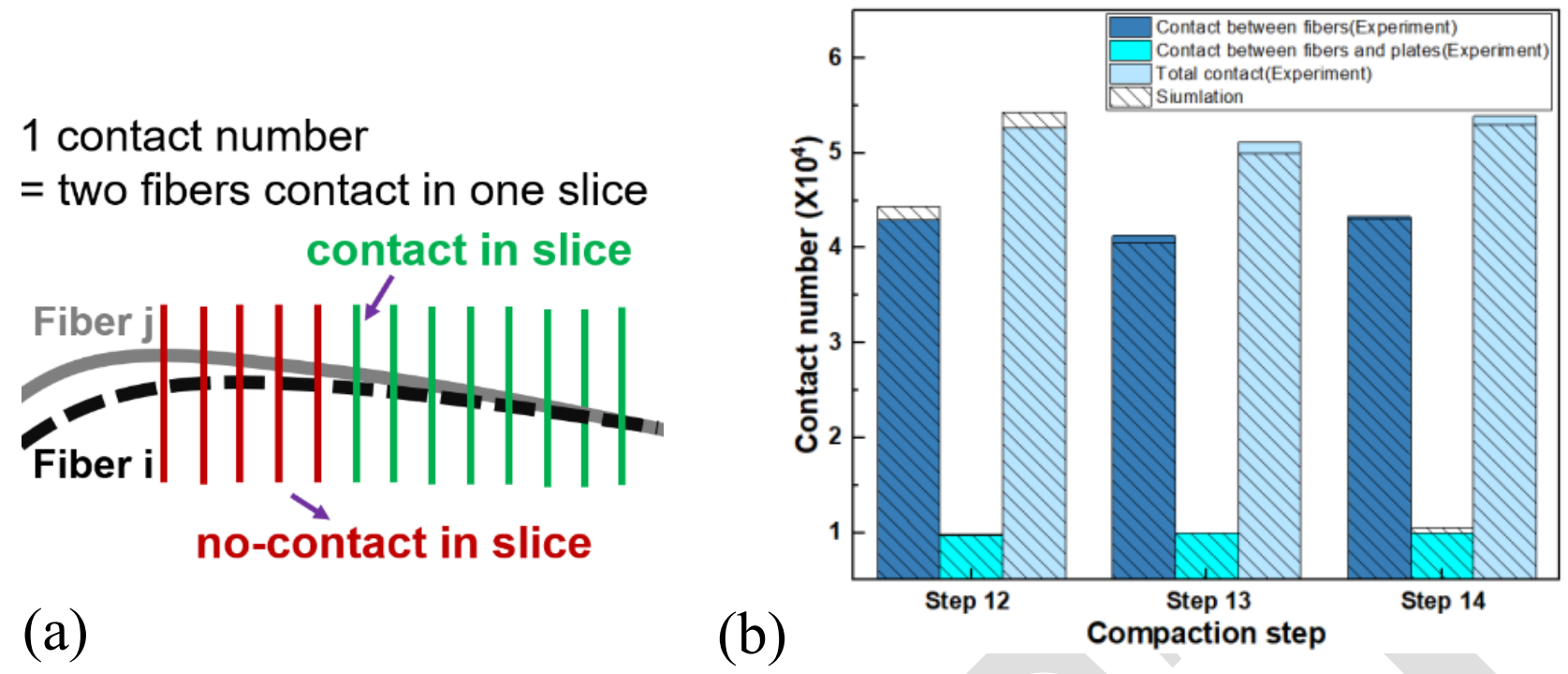

Figure 9. Contact number in the fiber bundle (a) Definition of contact number; (b) contact number in simulation and experiment compaction steps 12, 13, 14.

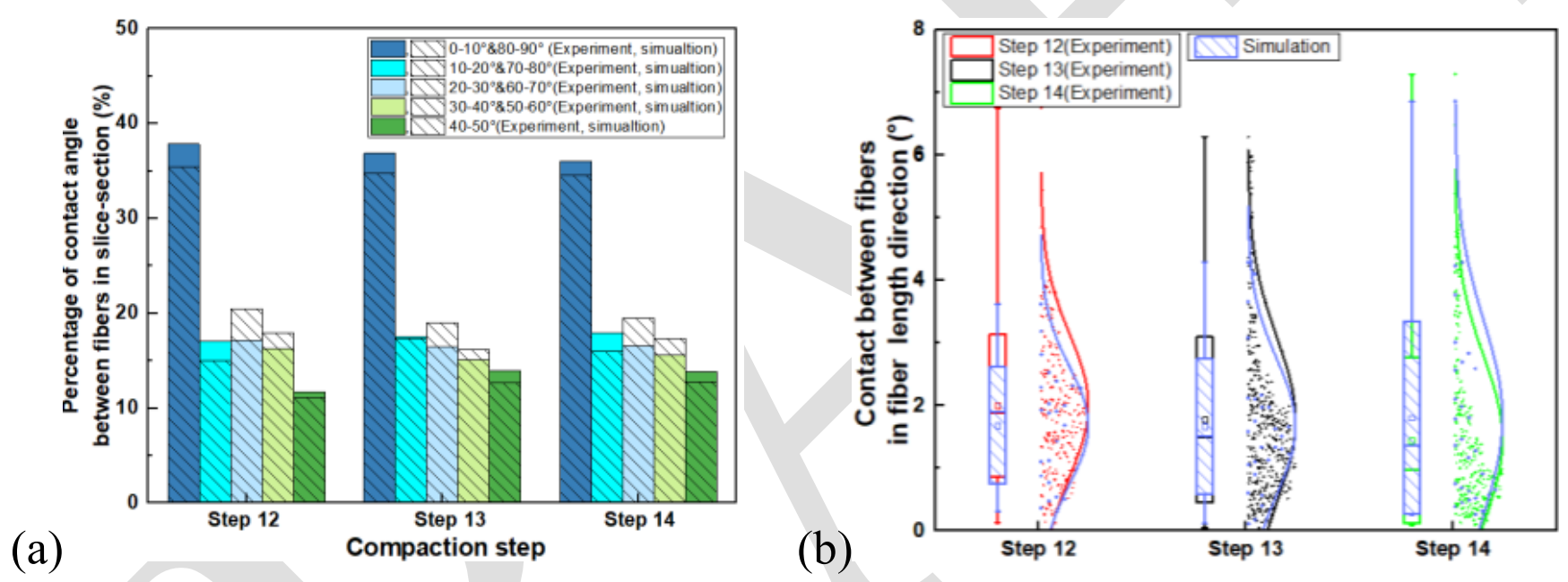

Figure 10. Contact angles (°) in sample 1 (a) between fibers in the fiber section $(\boldsymbol{\theta})$, (b) in the fiber length direction $(\boldsymbol{\gamma})$.

Fig. 11 shows the orientation distributions of the fibers in sample 1 with the compaction steps 12, 13, 14 shown in the unit spheres. The unit orientation tensor distribution of step 13 and step 14 in the $\boldsymbol{e_1 e_2}$ plane was almost the same, indicating that the main fiber rearrangement and void filling had already occurred by step 13. The fibers were largely stabilized in their configuration during step 14.

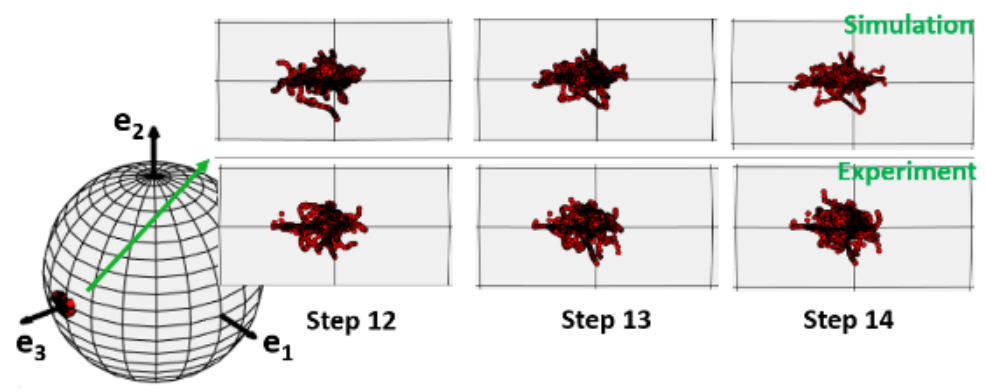

Figure 11. Orientation tensor of fibers with simulation and experiment compaction steps 12, 13, 14 (sample 1).

The boxplot of fiber crimp after the compaction steps 12, 13, 14 is shown in Fig. 12. Here again the consistency between the simulated and experimental values is noticeable. Although a slight decrease in crimp is observed at step 13, the global fiber waviness changes only marginally. Combined with the contact-angle distributions and the fiber-position analysis, this result indicates that the pressure drop at step 13 is not caused by a global structural reorganization. Rather, it is associated with a localized instability involving only one or a few fibers, as also suggested by the rearrangement visible in Fig. 8 (d, e).

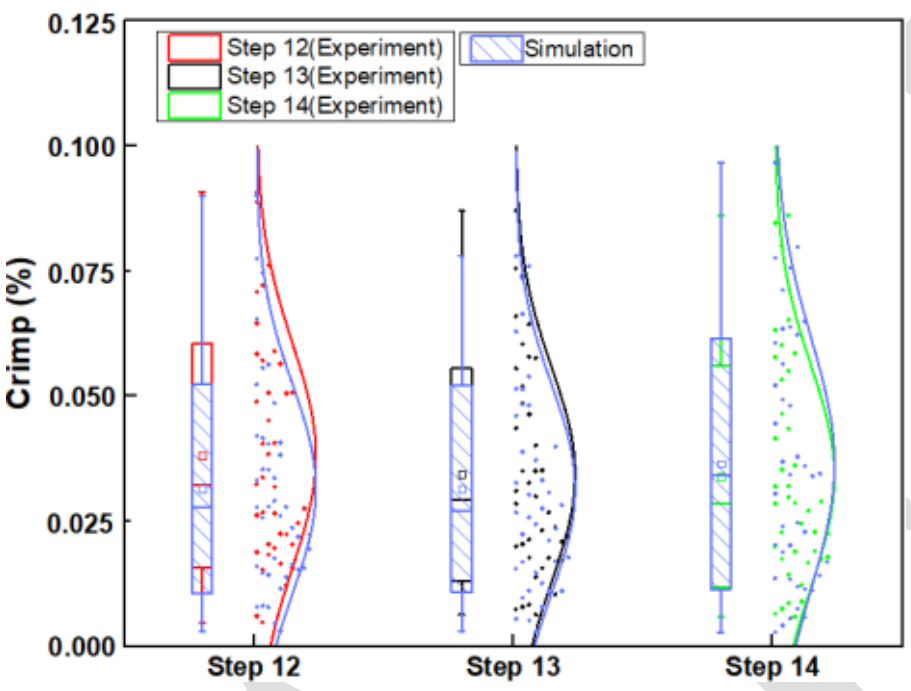

Figure 12. Boxplot of fiber crimp in the simulation and experiment compaction step 12, 13, 14 (sample 1).

The combined evolution of the microstructural descriptors suggests that the pressure drop observed around step 13 corresponds to a transient local instability in an almost jammed contact network. At this stage of compaction, the fibers are already highly confined, and only limited rearrangement remains possible. Under these conditions, a small local slip or bending event can induce a rapid redistribution of contact forces and produce a temporary decrease in the macroscopic compaction pressure. Several physical factors may contribute to this instability. First, the finite bending stiffness of the fibers allows local bending-assisted rearrangements once translational motion becomes highly constrained. Second, inter-fiber friction may promote intermittent stick-slip behavior, which can abruptly modify the contact locations and redistribute the load transfer paths. Third, the rigid confinement imposed by the compaction channel and plates may intensify local force concentration and facilitate localized rearrangement. Because the contact-angle distributions, orientation tensor, and crimp remain nearly unchanged, the pressure drop is better interpreted as a localized unlocking event than as a global reorganization of the whole bundle.

Overall, these results show that the proposed strategy is able to reproduce not only the global compaction response of an existing fiber bundle, but also the main features of its microstructural evolution. This validation supports the relevance of the numerical strategy developed here and motivates its use for subsequent parametric studies with virtual random fiber bundles.

### 3.3 Numerical random fiber bundles

The generator was developed to produce virtual fiber bundles with prescribed microstructural characteristics, including fiber number, misorientation, and undulation, thereby enabling systematic parametric studies of bundles with different internal architectures. In principle, a virtual bundle could be generated directly from CT-based in-situ tomography images by extracting fiber centerlines slice by slice and then connecting the corresponding points. However, when this strategy is applied directly to dense bundles, interpenetration between neighboring fibers becomes unavoidable once the fiber volume fraction exceeds approximately 45%. Conversely, preventing penetration by artificially increasing the spacing between fibers would lead to unrealistically large voids within the bundle. To overcome this limitation, each fiber was generated individually, and the complete bundle was subsequently assembled. As summarized in Fig. 13, the generation procedure combines geometrical construction in MATLAB with a gravity-driven settling stage in Abaqus®.

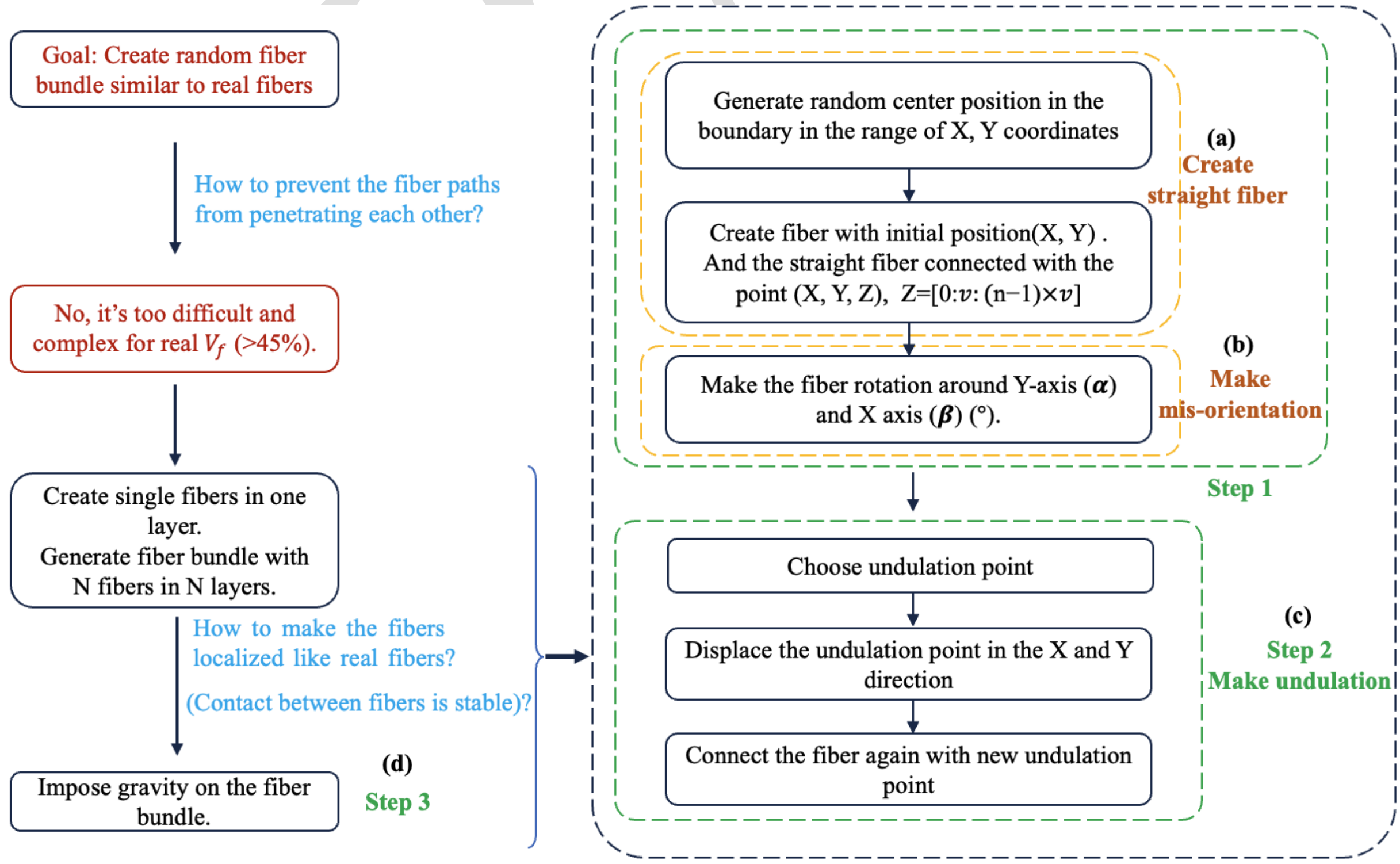

Figure 13. Flowchart of numerical generation step for random fiber bundle geometry.

Each fiber was first represented by a straight centerline discretized into $n$ slices, leading to a total length of $(n-1)\times v$ for $N$ fibers in Z-axis direction, where $v$ is the voxel size. The initial center point of each fiber was selected randomly within admissible bounds in the cross-sectional plane, shown in Fig. 14 (c), so that the entire three-dimensional geometry remained inside the compaction channel: For the X-th fiber, the Y-coordinate was positioned at $Y_N = \frac{1}{2}d + (N-1)d$, while the X -coordinate was chosen within the interval $X \in (r, X-2r)$, where $r$ denotes the fiber radius and X the width of the box. In this way, a set of initially straight and non-overlapping fibers was generated in separate layers. The random coordinates of the Nth fiber are $(X, Y)$, where $X \in (r, X-2r), Y = \frac{1}{2}d + (N-1)d$.

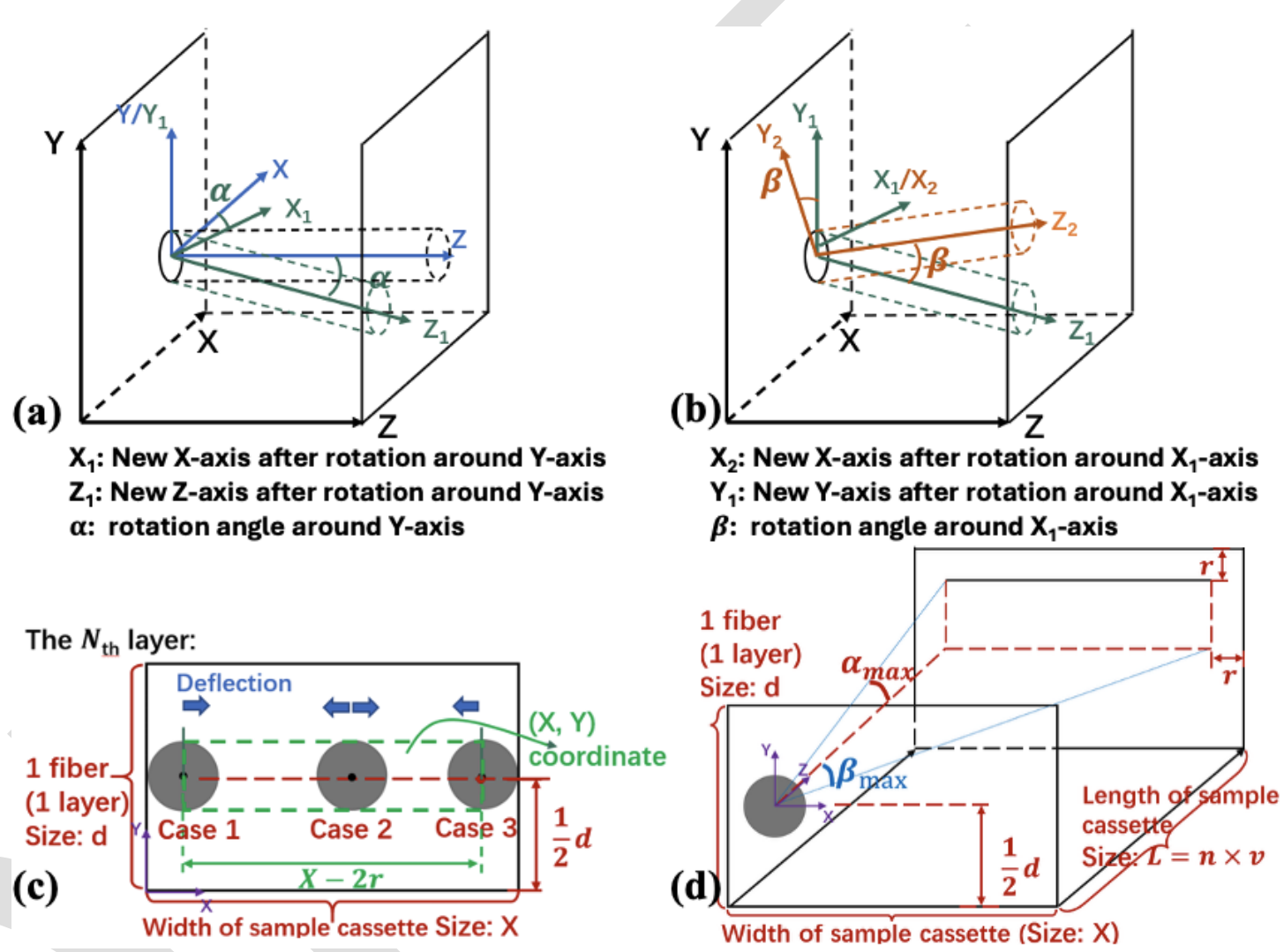

Figure 14. Principle for creating step 1 of Fig. 13. (a) Rotation of a single fiber in a separate layer around Y-axis ($\alpha$) (°). (b) Rotation of a single fiber in a separate layer around X-axis ($\beta$) (°). (c) Generation of random fiber initial position in step 1.1 (middle dot range) and three different cases for fiber deflection. (d) Scheme of maximum rotation angle $\alpha_{max}$ and $\beta_{max}$.

Fiber misorientation was then introduced through Euler rotations applied to the initial straight centerline (step (1. a)). As illustrated in Fig. 14 (a, b), the fiber was first rotated by an angle $\alpha$ about the Y-axis and then by an angle $\beta$ about the rotated $X_1$-axis, leading to the final $X_2$-$Y_2$-$Z_2$ coordinate system. The fiber orientation could thus be controlled through the parameters $\alpha$ and $\beta$. Twist was not considered in the present model, since it can reasonably be neglected for rovings. In addition, the admissible rotation angles were constrained by the dimensions of the compaction channel in order to prevent the fiber from leaving the cassette. Depending on the initial fiber position, the allowable deflection could therefore be restricted to one side or permitted on both sides, as shown in Fig. 14 (c, d):

$$|\alpha| \leq \alpha_{max} = \arctan\left(\frac{\frac{d}{2}-r}{L}\right) = \arctan(\frac{\frac{d}{2}-r}{n\times v}),\ |\beta| \leq \beta_{max} = \arctan\left(\frac{X-2r}{l}\right) = \arctan(\frac{X-2r}{n\times v})$$

Furthermore, it is important to ensure that the fiber bundle remains positioned inside the sample cassette inner wall (compaction channel). This concerns not only the initial position of the fibers (Fig. 14 (c)) but also to the positions of the deflected fibers. The fiber coordinates should therefore be contained within the boundaries of the sample cassette inner wall. Hence, it is essential to identify the randomly generated initial fiber position in Fig. 14 (c). For instance, the fiber can only deflect to one side (the right side) in case 1, whereas the fiber can deflect in both directions in case 2. Moreover, the angles $\alpha$ and $\beta$ should never exceed the maximum allowable values, as shown in Fig. 14 (d). Otherwise, fibers could exit the compaction support. After the fiber orientation had been prescribed, undulation was introduced in order to reproduce the non-straight morphology observed in real bundles. In the present section, undulation refers to the prescribed wavy geometrical feature imposed during fiber generation. It was created by inserting one or more control points along the fiber and displacing them in the longitudinal and transverse directions, as illustrated in Fig. 15. The fiber morphology after deflection of the fiber has been generated is shown in Fig. 15 (b). A bending point is inserted and, to illustrate this, an example is indicated in Fig. 15 (a). The final bending point is obtained by displacing the selected points (circle point) on the fiber in the Z, X, Y directions. The new curved fiber (dotted curved line) is obtained with one bending point. The strategy used to insert the bending point is the following: the first step consists in choosing the initial insertion point (circle point) so that the equipartition of the fiber length is ensured, as shown in Fig. 15 (b); then a displacement is applied on that point in the longitudinal direction (Z direction). To ensure a distributed placement of these bending points along the fiber length, the insertion slice was selected according to a Gaussian distribution centered at the middle of the fiber, with standard deviation $\sigma_{u_z}$ (Fig. 15 (c)). The selected point was then shifted in the $X_2$ and $Y_2$ directions using the statistical parameters $\mu_{u_X}$, $\sigma_{u_X}$, $\mu_{u_Y}$, and $\sigma_{u_Y}$, as shown in Fig. 15 (d). This procedure yielded curved fibers whose undulation level could be controlled through both the number of bending points and the amplitude of the prescribed displacements.

A direct assembly of such misoriented and undulated fibers would nevertheless lead to frequent intersections at high packing density. For this reason, the fibers were initially generated in separate regions so that their geometry could be prescribed independently without interpenetration (steps 1 and 2 in Fig.14). The complete bundle was then imported into Abaqus® (step 3), where a gravity-driven settling step was applied to bring the fibers into contact and to form a realistic bundle morphology. This final relaxation stage made it possible to obtain a dense fiber bundle with physically admissible contacts while avoiding artificial penetrations and preserving the targeted geometrical statistics as closely as possible.

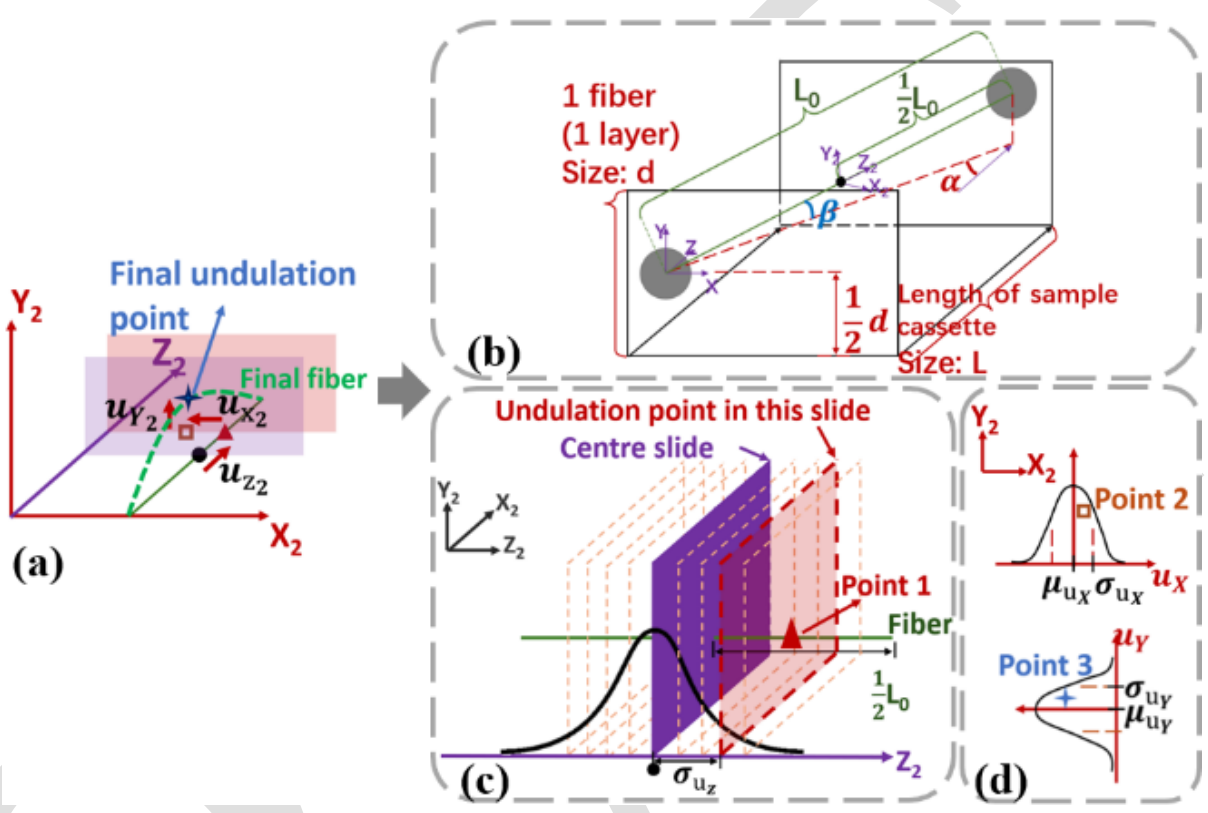

Figure 15. Principle of undulation generation by inserting a bending point (example with only one bending point). (a) Main step. (b-d) Detailed steps.

### 3.4 Validation of the generated numerical fiber bundles

In order to validate the ability of the virtual estimator to correctly represent the desired fiber bundle, 3 microstructures (geometries 1, 2 and 3) were generated with parameters extracted from sample 2 (Tab. 4). After imposing gravity, the $V_{f_0}$ of geometries 1 to 3 was 46.7%, 46.7% and 47.6% respectively. These values remain within the range of those obtained experimentally (about 50%).

The compaction test on these geometries was carried out and the results are compared to those of the experimental test on sample 2 (Fig. 16). As expected, since the microstructure parameters are similar, these three fiber bundle geometries showed pressure trends consistent with the experimental and the simulation reconstruction results. The compaction pressure decreased to around $V_f$ = 67%. Obviously, this fluctuation varies in magnitude as well as slightly in position ($V_f$) but the phenomenon remains. Even if the internal structure parameters of different fiber bundle models are basically the same, the positions of the fibers are different, resulting in different sliding between fibers at the late stage of compaction. This is due to the complex and random nature of the 3D structure of fiber bundles, in which even slight differences in structure can lead to significantly different deformation behaviors during compaction even if the initial structures are similar. In the early stage of compaction, the sliding of the fibers against each other accounts for the major part, making it easier to adjust the relative position between fibers during this stage. As a result, the compaction curve is basically the same. However, as compaction continues, relative movement between fibers becomes more difficult, and some fibers may bend or even twist, which in turn affects the stress distribution within the fiber bundle. This difference becomes particularly pronounced in the later stages of compaction, where small changes in compaction may lead to large stress or shape differences.

Table 4. Parameters of geometries 1-3 created by the virtual numerical random fiber generator

| Fiber number | 40 | Fiber radius (mm) | 0.25 | The size of the box (mm) | 5 |
|---|---|---|---|---|---|
| Voxel size (mm) | 0.009 | Slice number n | 1522 | The size of layer for each fiber (mm) | 2 |
| Rotation angle (°) | $\mu_\beta = 0.06,\ \sigma_\beta = 1.09$<br>$\mu_\alpha = -1.51,\ \sigma_\alpha = 2.29$ | | | Bending point number | 1 |
| X, Y, Z direction shift value (mm) | $\sigma_{u_Z} = 110$ slices; $\mu_{u_X} = 1.2,\ \sigma_{u_X} = 0.14;\ \mu_{u_Y} = 0.2,\ \sigma_{u_Y} = 0.2$ | | | | |

The reliability of the virtual numerical random fiber generator was verified in terms of the consistency between the experiment, the reconstruction simulation, and the created fiber bundle compaction curves. All these results validate the use of this virtual estimator to drive parametrical studies, opening a wide range of possibilities to better understand behavior of fiber bundles and go further in the identification of behavior laws.

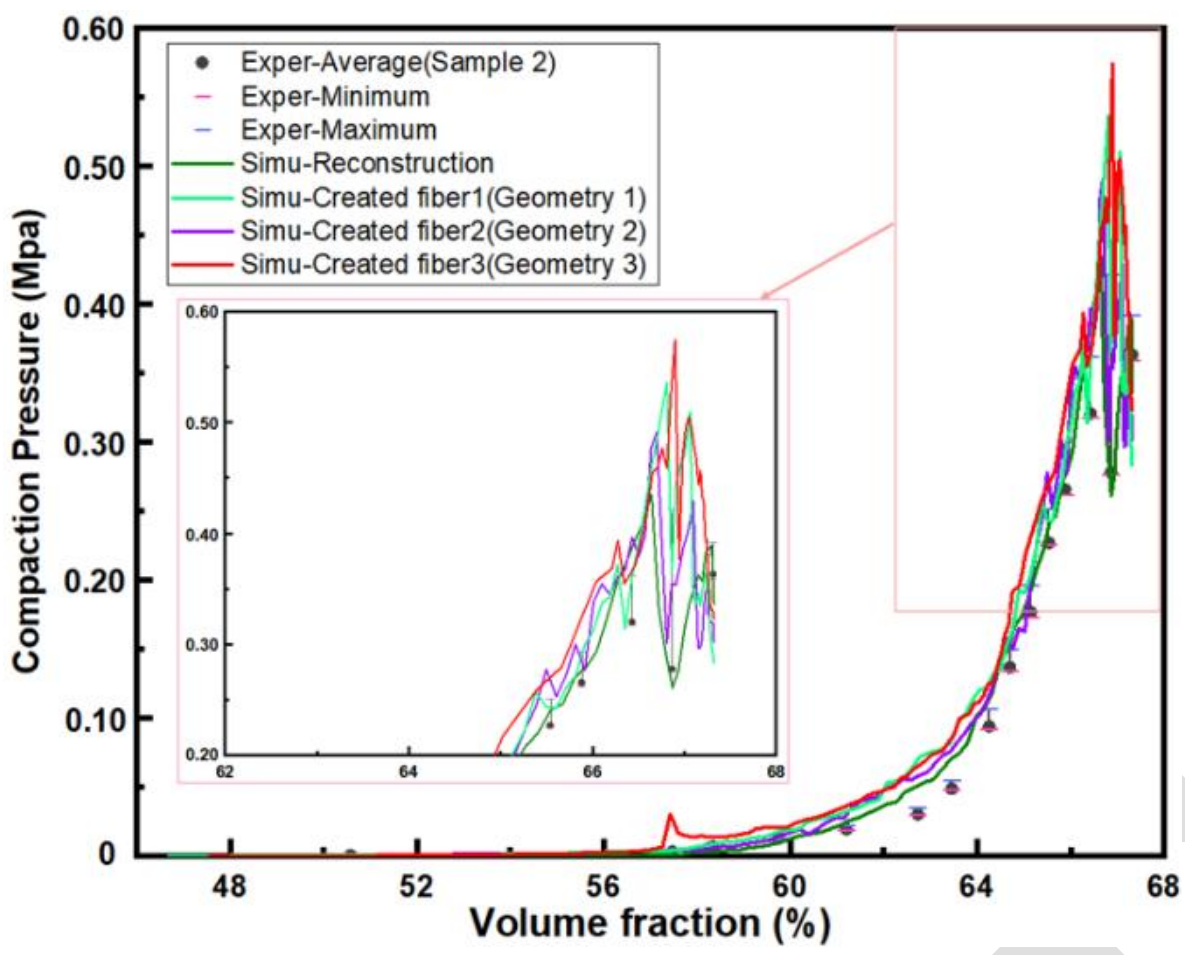

Figure 16. Compaction test with 3 numerical fiber bundle geometries, created by the virtual numerical random fiber bundle generator.

### 3.5 Effect of fiber undulation on compaction behavior

By setting different undulation parameters of the virtual numerical random fiber bundle generator, a series of compaction simulations on fiber bundles was performed with the aim of exploring the influence of the undulation of the fibers on the compaction performance. In order to study the effect of undulation of the fibers in the fiber bundle, 4 geometries of fiber bundles were created for compaction simulations and called geometries 4 to 7. The compaction curves were compared, as well as their microstructural features. Geometries 4 to 7 are all composed of 40 fibers; for all four geometries, the basic parameters are the same as in Tab. 4 (fiber diameter, mis-orientation, etc.). Besides the bending point number, the number of bending points increased from 1 to 4 for geometries 4 to 7 respectively, and $V_{f_0}$ was 46.7%, 43.5%, 34% and 27.8%, respectively.

Fig.17 shows the relationship between $V_f$ and compaction pressure. At first, only gravity is imposed on the initial geometries 4 to 7, created by the fiber bundle generator. For more undulated and entangled fiber bundles compacted to the same volume fraction, more pressure is required. The undulation and entanglement of the fibers increase the contact number or length, and thus, the interaction forces between the fibers. This prevents fibers from slipping and rearranging and thus requires a greater external pressure to achieve the same compaction. In addition, bent fibers may spatially form more locked structures that are difficult to break during compaction, thus also increasing the force required for compaction. This phenomenon highlights the importance of the structural intricacy within the fiber bundle on its compaction ability. It can be seen, especially for geometry 7 which is the most irregular one, that the increase in the compression stress leads to instabilities especially at high volume fractions, resulting in significant drops in the compaction curve.

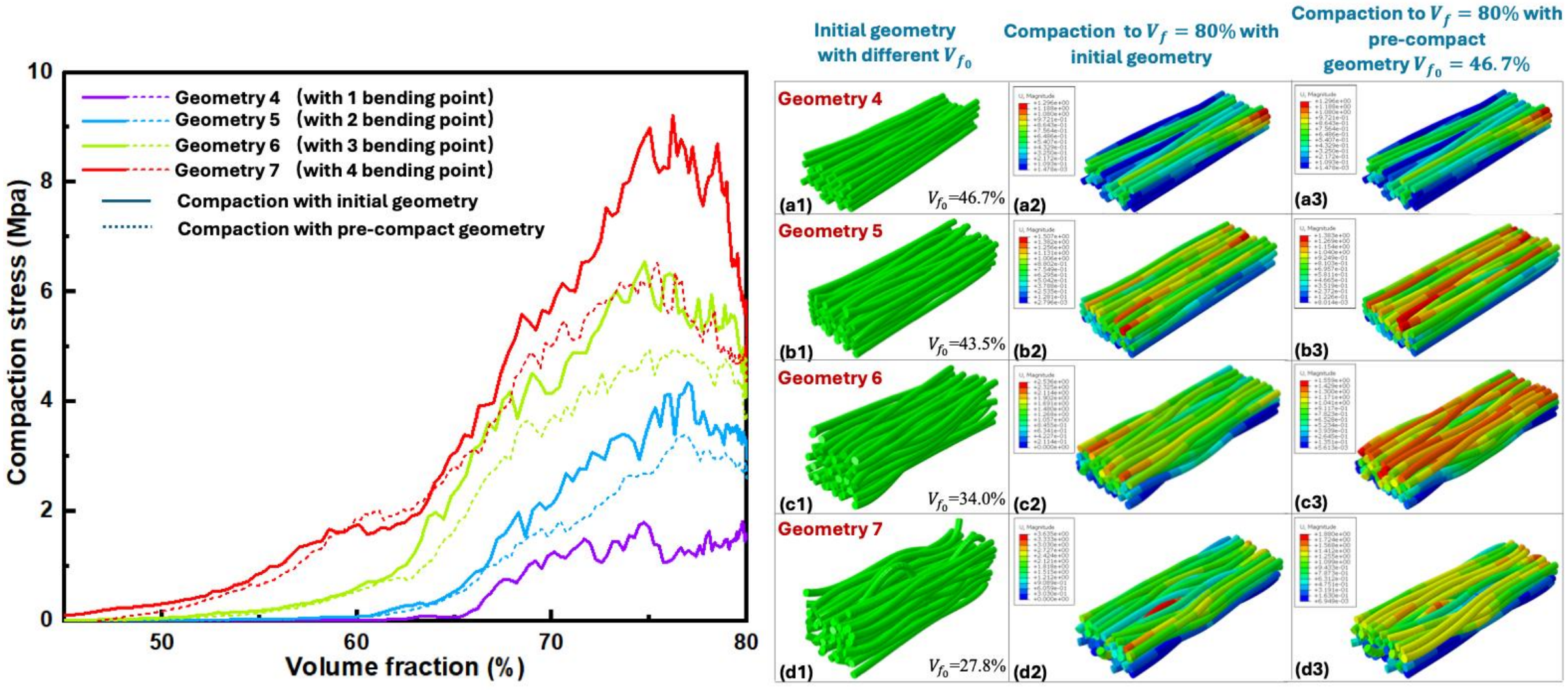

Figure 17 Relationship between Volume Fraction and Compaction Pressure: geometries 4 to 7, created with different undulation parameters. (a1-d1) fiber bundle geometries with bending point of 1-4. (a2-d2) fiber bundle 3D geometries after compaction to $V_f = 80\%$ with initial geometries. (a3-d3) fiber bundle 3D geometries after compaction to $V_f = 80\%$ with pre-compact geometry, which $V_{f_0} = 46.7\%$.

In order to exclude the effect of different initial volume fractions on compaction, geometries 4 to 7 were first compacted to $V_f$=46.7%, and then the compacted fiber bundle geometry model was re-imported into ABAQUS® for the second compaction simulation (Fig.17, dotted line). Compared to the compaction resulting from different initial fractions (Fig.17, solid line), the compaction pressure dropped at $V_f$=70 to 80%, and the final compaction pressure dropped by 0.8 Mpa, 0.4 Mpa and 0.3 Mpa respectively for geometries 5, 6 and 7. However, this does not change the conclusion concerning the influence of undulation on the compression behavior. During pre-compaction (compaction with low $V_{f_0}$), the fibers undergo a predominantly ordered rearrangement, and this rearrangement may reduce the friction and resistance between them during subsequent compaction. Moreover, pre-compaction results in fewer voids between fibers, and the fiber bundles become tight and more structurally stable. Thus, lower pressures are required to further increase the volume fraction. However, there appears to be a maximum volume fraction for compacted materials due to the locking of the geometry imposed by friction.

All samples, whether pre-compacted or not, may reach a similar state of structural locking, in which all the fibers are tightly packed, leaving few possible movements for the fibers. Thus, both pre-compacted and un-pre-compacted fiber bundles should exhibit a similar physical confinement, and thus the final compaction pressures for both are in the same range, consistently with Fig.17 (compaction curve) and Fig.17 (b2-c2 and b3-d3).

The microstructure indicators of geometries 4 to 7 were analyzed after the development of indicator tools, to understand the effect of undulation parameters on fiber bundle geometries before and after compaction. Fig. 18 shows the number of contacts between fibers, fibers and compaction plates and total contact number of geometries 4 to 7. The entangled fiber bundle geometries 4 to 7 have less total contact with each other in the uncompacted state. Undulation provides the fibers with more freedom for spatial deformation, which keeps the fibers widely spaced from each other. As a result, the initial volume fraction of bent and twisted fiber bundles is low for the same number of fibers. That is, the volume occupied in space is relatively large. When external pressure is applied to the fiber bundles, these fibers are more likely to be rearranged and densified during the deformation.

After compaction the trend is inverted, showing an increase in the contact numbers with undulation, as depicted in Fig. 18. This is logical because entanglement requires more fibers to be in contact in order to reach the same volume fraction. However, the small differences in the values after compaction also show that the number of contacts is certainly not the only parameter that drives the compression stiffness. This was rather unexpected, and needs to be kept in mind for the analysis of the compression behavior of fibrous media.

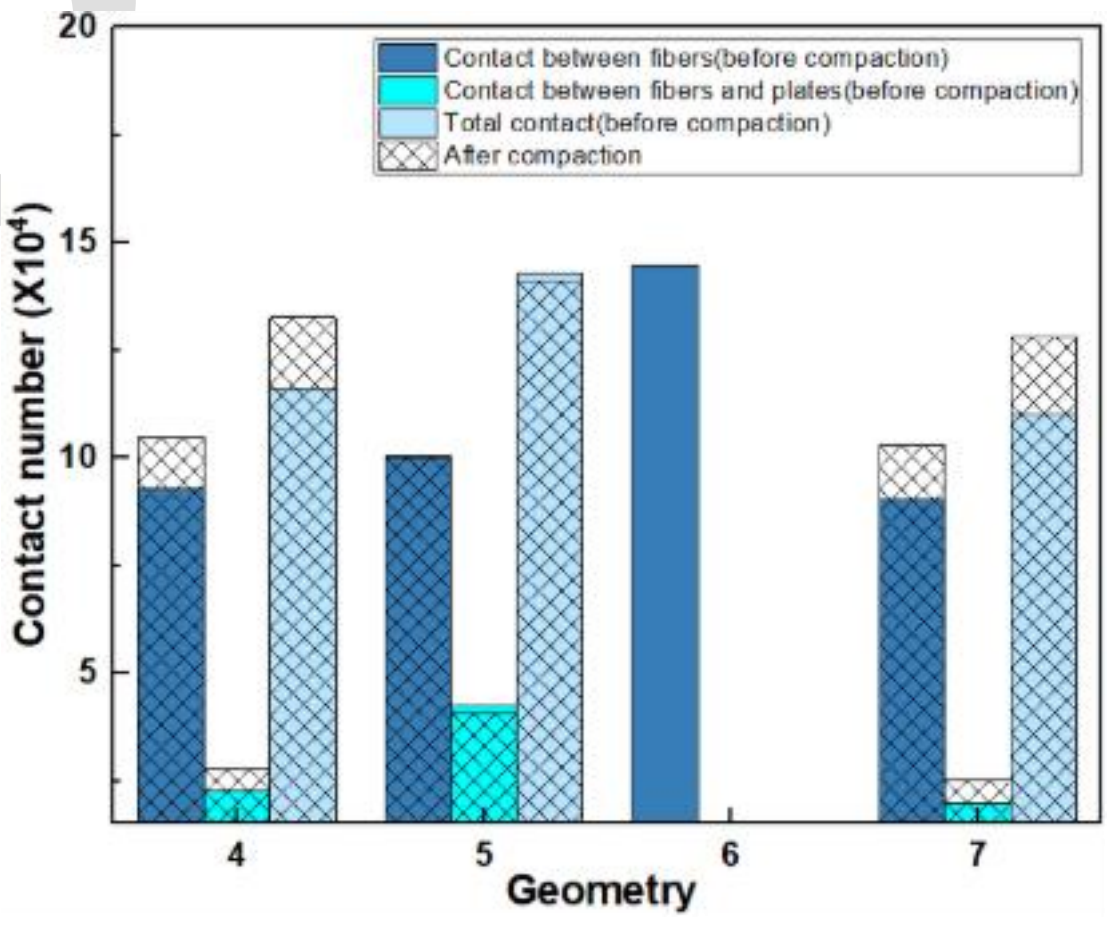

Figure 18. Contact number between fibers in the fiber bundle geometry

The orientation unit spheres and the orientation tensor $A_{33}$ are shown in Fig. 19. It should be noted that the $\Delta A_{33}$ (the difference in the orientation tensor $A_{33}$ before and after compaction) gradually increases with the increase in fiber bending points. This increase is related to the effect of fiber curvature on inter-fiber voids and interactions. Bent or entangled fibers need to be rearranged to fit the spatial constraints during compaction, a process that leads to more complex contact and force transfer between fibers, compacting the overall directionality of the fiber bundle. This effect is more pronounced when the number of fiber bending points increases, leading to more complex changes in compaction effects and fiber alignment directionality, which are reflected in changes in $A_{33}$. However, the reorganization cannot lead to the same final values; rather, it leads to a realignment and straightening of the fibers until locking of the geometry occurs. The more entangled the initial configuration is, the more entangled the locking configuration. The average and standard crimp values of fiber bundle geometries 4 to 7 are presented in Tab. 5, and the projected curvature radii in the XZ ($\overline{R_{XZ}}$) and YZ planes ($\overline{R_{YZ}}$) in Tab. 6. The results confirm the previous analysis.

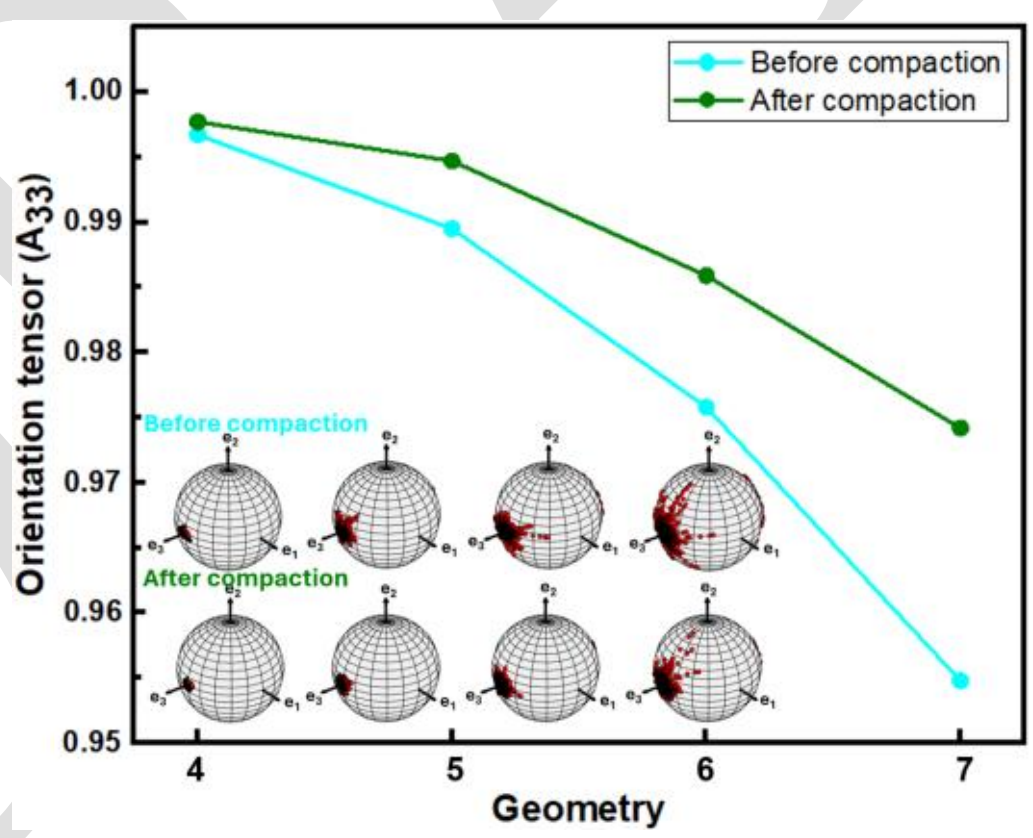

Figure 19 Orientation $\boldsymbol{A_{33}}$ and orientation unit spheres of Geometries 4 to 7 before and after compaction.

Table 5. Crimp of all the fibers in geometries 4 to 7 (%).

| Compact | $\mu$ | $\sigma$ | $\mu$ | $\sigma$ | $\mu$ | $\sigma$ | $\mu$ | $\sigma$ |
|---|---|---|---|---|---|---|---|---|
| | Geometry 4 | | Geometry 5 | | Geometry 6 | | Geometry 7 | |
| Before | 0.077 | 0.075 | 0.317 | 0.433 | 0.833 | 1.154 | 2.115 | 3.840 |
| After | 0.072 | 0.006 | 0.202 | 0.170 | 0.570 | 0.449 | 1.147 | 1.238 |

Table 6. Projected curvature radii $\overline{\boldsymbol{R_{XZ}}}$ and $\overline{\boldsymbol{R_{YZ}}}$ ($\mathbf{mm}$).

| Compact | $\overline{R_{XZ}}$ | | $\overline{R_{YZ}}$ | | $\overline{R_{XZ}}$ | | $\overline{R_{YZ}}$ | | $\overline{R_{XZ}}$ | | $\overline{R_{YZ}}$ | | $\overline{R_{XZ}}$ | | $\overline{R_{YZ}}$ | |
|---|---|---|---|---|---|---|---|---|---|---|---|---|---|---|---|---|
| | $\mu$ | $\sigma$ | $\mu$ | $\sigma$ | $\mu$ | $\sigma$ | $\mu$ | $\sigma$ | $\mu$ | $\sigma$ | $\mu$ | $\sigma$ | $\mu$ | $\sigma$ | $\mu$ | $\sigma$ |
| | Geometry 4 | | | | Geometry 5 | | | | Geometry 6 | | | | Geometry 7 | | | |
| Before | 139 | 19 | 187 | 30 | 61 | 38 | 74 | 42 | 35 | 29 | 46 | 38 | 21 | 13 | 21 | 14 |
| After | 240 | 18 | 280 | 66 | 50 | 42 | 86 | 106 | 27 | 13 | 54 | 54 | 17 | 14 | 31 | 41 |

The crimp of the fibers in geometries 4 to 7 increases, which is accompanied by an increase in the number of bending points. This phenomenon existed before and after compaction, and is related to the previously mentioned entanglement within the fiber bundle. Fibers partially straighten during compaction. However, the more disordered the locking configuration imposed by entanglement is, the less they straighten. This is the reason why fibers remain curved, and thus the crimp is higher and the projected radii are smaller. Combining the compaction curve (Fig.17) and the crimp of the fiber bundle geometries, the geometry with high crimp requires more pressure to rearrange and movement to achieve the same volume fraction as fibers with low crimp.

Analysis of the radii sheds interesting light on the fiber transformation. As shown in Tab. 6, before compaction simulation, the $\overline{R_{XZ}}$ and $\overline{R_{YZ}}$ of geometries 4 to 7 gradually decrease, consistently with the undulation from geometry 4 to 7. After compaction, the difference remains and in order to achieve the targeted $V_f$, the curvature radius $\overline{R_{YZ}}$ shows that the fibers become straighter in the YZ plane, because the compaction force acts on the fibers, forcing them to adjust their shape to take up less space. Fibers tend to align in the direction of fiber length, so crimped fibers will partially straighten under pressure. However, due to the initial microstructure undulation properties (entanglement structure) of the fibers in geometries 5 to 7, the fibers are interlocked, they cannot straighten and so retain a certain curved structure. One can also notice that there is a small decrease in $\overline{R_{XZ}}$ for the undulated configuration, again showing the ability of the fibers to bend locally to occupy gaps.

These results demonstrate the influence of fiber undulation on the compaction behavior, even for the same volume fraction, which indicates that neither the volume fraction nor the contact number alone drive the compaction behavior. The increase in undulation leads to an increase in the compression stiffness, within the range of study, which was the sought-after answer. This illustrates that the virtual estimator makes it possible to tackle the influence of a parameter on the mechanical response, here, compression. This description reflects the microscopic behavior and structural changes of fiber bundles during compaction, during which the interaction between fibers involves rearrangement and also mechanical interactions. Undulated fibers, because of their complex shape, lead to the formation of a tangled structure between fibers, a structure that increases compaction resistance.

The final compaction strength of both pre-compacted and un-pre-compacted fiber bundles converged after compaction to the same volume fraction. This indicates that the structure of the fiber bundle tends to reach a steady state during compaction, and the close contact between fibers makes further deformation or rearrangement difficult.

**4 Conclusion**

The goal of the study takes place in the understanding of fiber bundles behavior and the future implementation of a mesoscopic behavior law. The strategy is to implement a microscopic approach to be able to easily drive parametric studies.

1) A numerical strategy was first developed from a real fiber bundle and validated against confined compression experiments combined with X-ray microtomography. The superposition of experimental and numerical microstructures showed good agreement, with a fiber-position error of 5.2%. The numerical load–volume fraction curves also reproduced the experimental response, confirming the validity of the simulation strategy.

2) This validation further allowed several microstructural descriptors, including contact number, contact angle, fiber orientation and crimp, to be defined and compared. Their consistency between real and numerical bundles shows that the model captures not only the global compaction behavior but also the main microstructural evolution during deformation. These descriptors were then used as input parameters for a virtual fiber-bundle generator.

3) Based on the effective simulation strategy, a second need was to create a numerical fiber bundle generator, which controls fiber misorientation and waviness at the single-fiber level by introducing 1-4 bending points, followed by gravity-driven settling to create realistic contacts without fiber penetration. Generated bundles based on the real microstructural parameters reproduced the experimental compaction range. Parametric results showed that increasing waviness strengthens inter-fiber interactions, increases transverse stiffness and requires a higher load to reach the same fiber volume fraction.

4) The proposed framework is mainly applicable to dry quasi-parallel bundles under quasi-static transverse compaction. Its explicit fiber-scale contact description provides detailed mechanical insight but limits direct use in large textile-scale simulations. In the present computational setting, one simulation required about 30 min, which remains suitable for offline parametric studies and for building reduced mesoscopic laws. Future work will extend the approach to glass and carbon fibers, different bundle architectures, multiscale coupling and more complex loading paths, including bending, shear, cyclic and dynamic loading.

**Acknowledgments**

This project was supported by the China Scholarship Council (CSC), the project of "National constructed high-level university-sponsored graduate programs" (Funding number: 202008120116).

**Ethics declaration**

Not applicable.

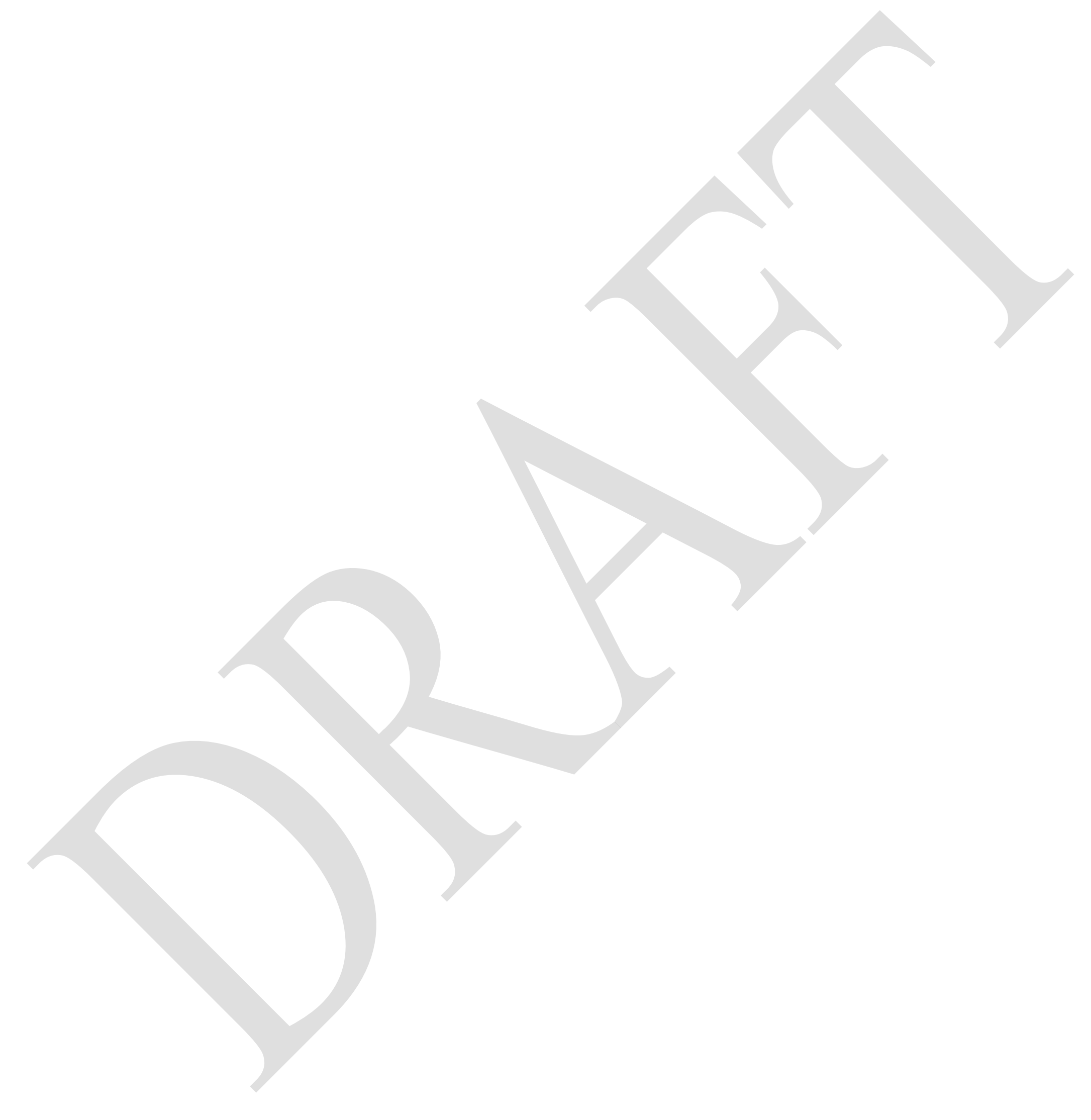